\documentclass[preprint2]{aastex}

\shorttitle{unpulsed emission from pulsars}
\shortauthors{Basu, Athreya \& Mitra}

\begin{document}


\title{Detection of off-pulse emission from PSR B0525+21 and PSR B2045--16}


\author{Rahul Basu}
\affil{National Centre for Radio Astrophysics, P. O. Bag 3, Pune University
   Campus, Pune: 411 007. India}
\email{rbasu@ncra.tifr.res.in}

\author{Ramana Athreya}
\affil{Indian Institute of Science Education Research (IISER) - Pune\\900, NCL Innovation Park, Homi Bhabha Road, Pune: 411 008. India}
\email{rathreya@iiserpune.ac.in}

\and

\author{Dipanjan Mitra\altaffilmark{1}}
\affil{National Centre for Radio Astrophysics, P. O. Bag 3, Pune University
   Campus, Pune: 411 007. India}
\email{dmitra@ncra.tifr.res.in} 


\altaffiltext{1}{Part of the work was done while at NAICenter, Arecibo 
Observatory}


\begin{abstract}
We report the detection of off-pulse emission from two long period pulsars 
PSR B0525+21 \& PSR B2045--16 using the Giant Metrewave Radio Telescope (GMRT).
We recorded high time resolution interferometric data at 325 MHz and gated 
the data offline to separate the on- and off-pulse sections of the 
pulsar. On imaging the two temporal sections separately, we detected a point source in 
the off-pulse images of both the pulsars coincident with the pulsar locations 
in the on-pulse images. The long periods (3.75 and 1.96 s respectively) 
and low spin-down energies ($\dot{E} \sim 10^{31} $ erg~s$^{-1}$) of the two 
pulsars argue against a pulsar wind nebula (PWN) origin for the off-pulse
emission. The derived estimates of the ISM particle density required to drive 
a PWN do not support such an interpretation. A magnetospheric origin for 
the off-pulse emission raises questions regarding the location of the emission
region, which is an important input into pulsar emission models.
 
\end{abstract}





\section{\large Introduction}

We report the detection of off-pulse emission from two long period pulsars 
PSR B0525+21 and PSR B2045--16 using the interferometric mode of the Giant 
Metrewave Radio Telescope (GMRT) in the 325 MHz frequency band. The 
``main-pulse'' (or the equivalent term ``on-pulse'' used in this paper) is the 
emission from within the polar-cap region of a pulsar; and the off-pulse is the
emission outside this main-pulse. Off-pulse emission from pulsars has been a 
subject of interest since the discovery of pulsars four decades ago.

According to the rotating vector model (RVM) proposed by Radhakrishnan 
\& Cooke (1969) the observed pulses are due to relativistically beamed 
radiation along the open dipolar field lines. The plane of linear 
polarization traces the magnetic field line associated with the emission 
at every instant. Several statistical studies have revealed that pulsar 
radio emission is in the form of a circular beam arising from the polar 
cap (i.e. the main-pulse) with an opening angle $\rho$, and from a height 
above the magnetic pole equal to 1--2 per cent of the radius of 
the light cylinder. The width of the observed pulse is a geometrical function 
of $\rho$ (which itself is a function of the emission height), the pulsar 
period, the angle between the rotation axis and the magnetic axis ($\alpha$) 
and the angle between the magnetic axis and the line of sight to the observer
($\beta$). The angles $\alpha$ and $\beta$ can be estimated by fitting 
the RVM to the swing of the polarisation position angle (PPA) across the 
pulse \citep{evert01, mitra04}.

The observed main-pulse covers 5-10 per cent of the pulse period for 90 per 
cent of the pulsars. In several pulsars this main-pulse consists of multiple 
components including low-level bridge emission between such components 
(Rathnasree \& Rankin 1995). The pulse width scales as 1/sin($\alpha$), from 
the geometry of the emission model; in the rare case of a pulsar with closely 
aligned magnetic and rotation axes (i.e. $\alpha \sim 0\degr$ or 180\degr) 
the pulse width can 
be as high as 100 per cent. Several pulsars have low level but detectable 
bridge emission between two widely spaced pulse components, making them 
candidates for an aligned rotator geometry. In rare cases when $\alpha$ is 
close to 90\degr, emission from the opposite pole (i.e. the inter-pulse) 
may be observed at 180\degr phase from the main-pulse. In all these instances 
the observed emission is believed to lie within the polar cap (i.e. main-pulse) 
as described earlier. 

In a few pulsars, low level emission components known as pre-/post-cursors (PPC)
have been observed outside the main-pulse (e.g. see Mitra \& Rankin 2010 for a 
discussion). In these pulsars, all geometrical evidence indicate that the main 
pulse is consistent with emission from open field lines. The 
PPC components appear highly polarized and are far from the 
main-pulse. The discovery of the PPC component about 60\degr away from the main 
pulse in PSR B0943+10 is particularly interesting, as the line of sight almost 
grazes the emission cone for this pulsar \citep{backus10}. Hence the PPC 
emission originates either from a much larger height, where due to spreading 
of dipolar field lines the PPC component can lie far away from the main-pulse, 
or from the regions of closed field lines. Some pulsars occasionally emit giant
pulses which are believed to arise close to the light cylinder rather than the 
polar cap. Interestingly PPC pulsars have low spin-down 
luminosities ($\dot{E} < 10^{34}$~erg~s$^{-1}$), whereas the giant pulses arise
from pulsars with $\dot{E} \gtrsim 10^{34}$~erg~s$^{-1}$. The PPC components and
giant pulses are potential sources of magnetospheric off-pulse emission in 
pulsars. They challenge the conventional wisdom of pulsar radio emission 
arising only from open magnetic field lines close to the pulsar polar 
cap and raise questions about the origin of these emission components.

The other possible source of off-pulse emission is the interstellar medium
(ISM) around the pulsar. The pulsar loses most of its rotational energy in 
the form of a relativistic wind which, when confined by the surroundings, may 
form a pulsar wind nebula (PWN). Several types of PWNe are formed depending
on the confinement mechanism. Young and energetic pulsars are often located 
in their associated supernova remnant (SNR). The pulsar wind streaming into 
the ambient medium produces standing shocks resulting in a plerionic PWN (like 
the one observed in the Crab nebula). In older pulsars, where the surrounding 
SNR is likely to be dissipated, the relativistic particles may interact with 
the ISM magnetic field and radiate, creating ghost remnants (proposed by 
Blandford et al. 1973, but not observed till date). Finally, a pulsar moving 
through the ISM with supersonic speed can produce a bow-shock nebula, where 
ram pressure balance is established between the pulsar wind and the ambient 
medium. Several such bow shock nebulae have been detected in H$\alpha$, and 
some in the radio as plerionic bow shocks, but never in both (see Chatterjee 
\& Cordes 2002 for a detailed study). 

Several searches for PWNe have been conducted in the past with varying 
sensitivities and resolutions. However only in about 10 young pulsars have 
radio emission been detected outside the main-pulse and all of these are 
believed to be associated with PWNe. In all such cases the associated pulsars 
are young (10$^3$-10$^5$yr), have high spin-down luminosities ($\dot{E} 
\gtrsim 10^{35}$~erg~s$^{-1}$) and all except one are associated with SNRs 
\citep{gaens98, gaens00, stap99}. It is believed that the pulsar 
wind and the environment change as pulsars slow down and age, making them less
likely to harbour PWNe.

The present study was designed to detect radio emission away from the polar 
caps and not associated with PWNe. Therefore we targeted old and less 
energetic pulsars unlikely to harbour PWNe. Their profiles did not 
show any components outside the main-pulse. The targets were neither aligned
rotators nor were inter-pulsars. We attempted to detect emission within a pulse phase range of approximately 80-250\degr\ from the peak in the main-pulse.

We describe the gated interferometric observations and the data analysis 
in section 2, the results in section 3, the additional tests we carried out to 
confirm that the off-pulse emission were not instrumental artifacts in section 
4, and discuss the implications of the detections in section 5.

\section{\large Observations \& Analysis}


\subsection{Target Pulsars}

We chose two pulsars, B0525+21 and B2045--16, whose pulse 
profiles did not show any features outside the main-pulse \citep{gould98} and
were stronger than 50 mJy to be able to put a non-detection 
upper limit of at most 5 per cent of the pulsed flux. The properties of the 
targets are listed in table \ref{PulsProp}. We targeted long period pulsars
because:\\ 
1. PWNe are so far only known from short-period, energetic pulsars in the 
vicinity of SNR. A long period selects against both high energy 
pulsars and SNR association, and hence PWNe.\\
2. We wanted at least 8 bins (of 131 or 262 ms each) across the pulse 
period to cleanly separate the off-pulse and the on-pulse regions.

The temporal broadening across the 16 MHz bandwidth due to dispersion by the 
ISM was 183.3 ms for PSR B0525+21 (binwidth 262 ms, gate width 1.31 s), 
and 47.8 ms for PSR B2045--16 (binwidth 131 ms, gate width 0.655 s). 
Therefore, we did not have to dedisperse the signal.

\begin{table*}
\begin{center}
\caption{Properties of Pulsars selected for off-pulse studies.
\label{PulsProp}}
\begin{tabular}{lccccccc}
\tableline\tableline
Pulsar&period&     DM    &dist&$\tau_{c}$& $\dot{E}$  &${V_{trans}}$&$Flux_{325}$ \\
      & s  &$cm^{-3}pc$& kpc&   year   &erg~$s^{-1}$& km~$s^{-1}$ &   mJy \\
\tableline
B0525+21&  3.7455 &  50.937  &  2.28  & 1.48$\times10^{6}$ & 3.0$\times10^{31}$ &  229 & 80.5 \\
B2045--16& 1.9616 &  11.456  &  0.95  & 2.84$\times10^{6}$ & 5.7$\times10^{31}$ &  511 & 169.3 \\
\tableline
\end{tabular}
\tablenotetext{~}{\small Period, dispersion measure and proper motion 
(for $V_{trans}$) are from \citep{hobbs04}; the others are from 
\citep{taylor93}. $Flux_{325}$ is the expected flux at 325 MHz calculated from
values in \citep{lor95}.}
\end{center}
\end{table*}

\subsection{Interferometric imaging of the pulsars with the GMRT}

Interferometric observations are better than single dish measurements for this 
study for several reasons:\\
1. An imaging interferometer is only insensitive to constant flux density 
background along the spatial axes while a standard pulsar receiver is usually 
insensitive to the constant background along the time axis; the detection of 
off-pulse emission is essentially an attempt to find such a constant background
along the time axis.\\
2. Self-calibration of interferometric data can correct instrumental and 
atmospheric gain fluctuations on very short time scales. The corrections
are determined by the flux densities of the constant and bright background 
sources in the field and hence would not be affected by the pulse variation 
of the relatively weak pulsar.\\
3. The higher spatial resolution of an interferometer greatly reduces the 
coincidence of unrelated sources; thereby reducing the probability that the 
off-pulse emission is from an unrelated source within the synthesized beam.

We imaged the pulsars with the Giant Metrewave Radio Telescope (GMRT), an
aperture-synthesis radio interferometer located near Pune, India
\citep{swarup91}. The 30 antennas of 45 m diameter provide a maximum baseline
of 27 km and can be operated at 6 frequency bands between 50 and 1450 MHz. We
observed the pulsars at 325 MHz with a 16 MHz bandwidth split into 128
channels. The frequency was chosen for its optimal combination of resolution
(10 \arcsec) and sensitivity (few 100 $\mu$Jy in 4 hours of observing). The
high frequency-resolution was useful in flagging narrow-band radio-frequency
interference (RFI) and to avoid bandwidth smearing. The shortest integration
output by the hardware correlator is 0.131 s, though this data is usually
averaged and median filtered to provide a standard data output integration of
16.77s.

The two pulsars were first observed on 19 May, 2009 with 0.262s integration 
to keep the data output rate and size manageable. This time resolution 
(of 7 bins) was found to be inadequate in the case of PSR B2045--16 while 
the data was of poor quality ($>10$ missing antennas) for both targets. 
We reobserved the sources on 19 January, 2010, with integrations of 
0.131s for PSR B2045--16 (period 1.9616s), and 0.262s for PSR B0525+21 
(period 3.7455s). This allowed us to divide both the pulsar periods into 
14 time bins.

\subsection{Data Analysis}

Several earlier studies used the technique of online gating to separate the
on- and off-pulse regions using the timing information of pulsars, and 
accumulated time-averaged on- and off-pulse data \citep{strom90, stap99}. 
This resulted in relatively smaller size of data files and hence shorter 
computation times which was a major consideration 1-2 decades ago. The 
disadvantage was the impossibility of rectifying any errors during the gating 
process (see results and discussions in Gaensler et al. 2000). We recorded the
data at sufficiently high resolution and applied offline gating.

\subsubsection{Folding \& Gating}

In the absence of accurate absolute time reference in the interferometric 
mode we used offline gating to isolate the off-pulse region of the pulsar 
period. The 30 antennas of GMRT provide 30 self, and 435 cross visibilities 
at each integration interval (0.131s or 0.262s). We folded the self-data to 
obtain the pulsar profile and identify the on-and off-pulse regions. This was
used to gate the cross-antenna visibilities into on- and off-pulse data sets.

\begin{figure*}
\includegraphics[angle=0,scale=.62]{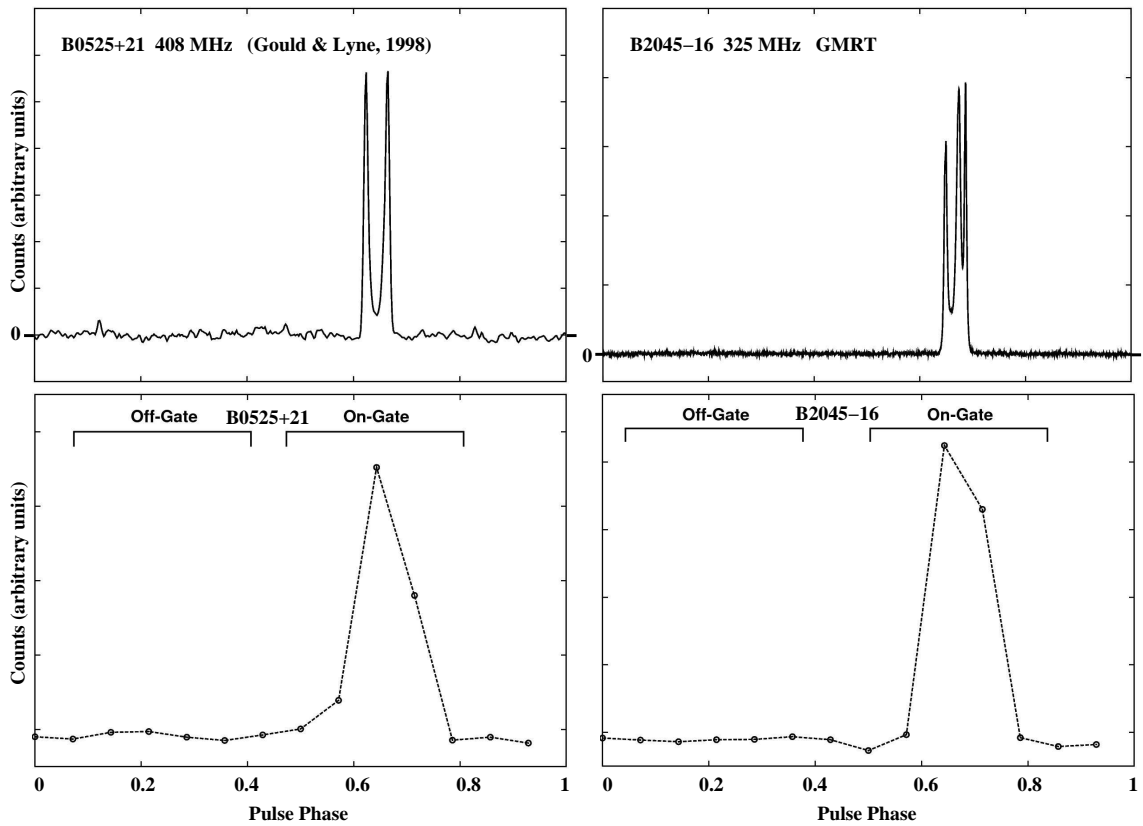}
\caption{Pulse profiles for our targets from standard pulsar mode high time
resolution observations (top row) and our folded interferometric self data 
(bottom row).
\label{fig_profileP}}
\end{figure*}

The temporal and frequency gain variations in the self-data were corrected by
normalising the instantaneous values by the local median. The timescale for the
temporal median was 30-50 times the pulsar period to ensure that the pulse
variation was retained. The data from all frequency channels and antennas were
averaged with robust sigma-clipping and processed using the standard folding
algorithm for pulsars (Hankins \& Rickett 1975) to determine the pulse profile.
The pulsar profiles and the off- and on-pulse gated are shown in Figure 
\ref{fig_profileP}.

The off- and on-pulse gates of PSR B0525+21 were centred on phases 0.24 and 
0.64, respectively. The 5 bins nearest to each were averaged to construct the 
off- and on-pulse data sets. The corresponding phases for PSR B2045--16 were 
0.21 and 0.67, respectively. The folding and gating procedures were carried out
using software developed for this purpose.

\subsubsection{Imaging}

The on- and off-pulse data were cleaned of radio frequency interference (RFI)
using the RfiX algorithm \citep{athreya09}. They were separately calibrated,
flagged and imaged in a standard manner using the NRAO AIPS package. The
flux-density scale of the images was determined from observations of the
calibrator 3C48 and the measurements of Baars et al.(1977) with the latest
corrections of Perley et al (1999; in AIPS). The data sets were taken through
several rounds of phase self-calibration followed by a final round of amplitude
and phase self-calibration. 

{\bf PSR B0525+21:} Observations of the calibrator 0521+166 were interspersed 
with the target to correct for the amplitude and phase gain fluctuations. The 
presence of the extremely strong and extended crab nebula (flux $>$ 1000 Jy 
and angular size $\sim$ 10\arcmin) at the edge of the primary beam (1.5\degr\
away) resulted in enhanced noise and strong ripples in the initial image.
Therefore, subsequent analyses were carried out with a lower UV cut-off at 
1.5 k$\lambda$ which reduced the artifacts and noise. This eliminated all 
structures larger than 2.3\arcmin\ from the image; this did not affect our 
purpose as our target source was expected to be much smaller than 2\arcmin. 

{\bf PSR B2045--16:} The calibrator 2137--207 was observed for amplitude and 
phase calibration. A pointing error resulted in the pulsar being located 
35\arcmin\ away from the field centre (60 per cent gain level of the primary 
beam). The primary beam correction was applied to the image to obtain the 
correct flux density for the pulsar.

\begin{figure}
\includegraphics[angle=-90,scale=.40]{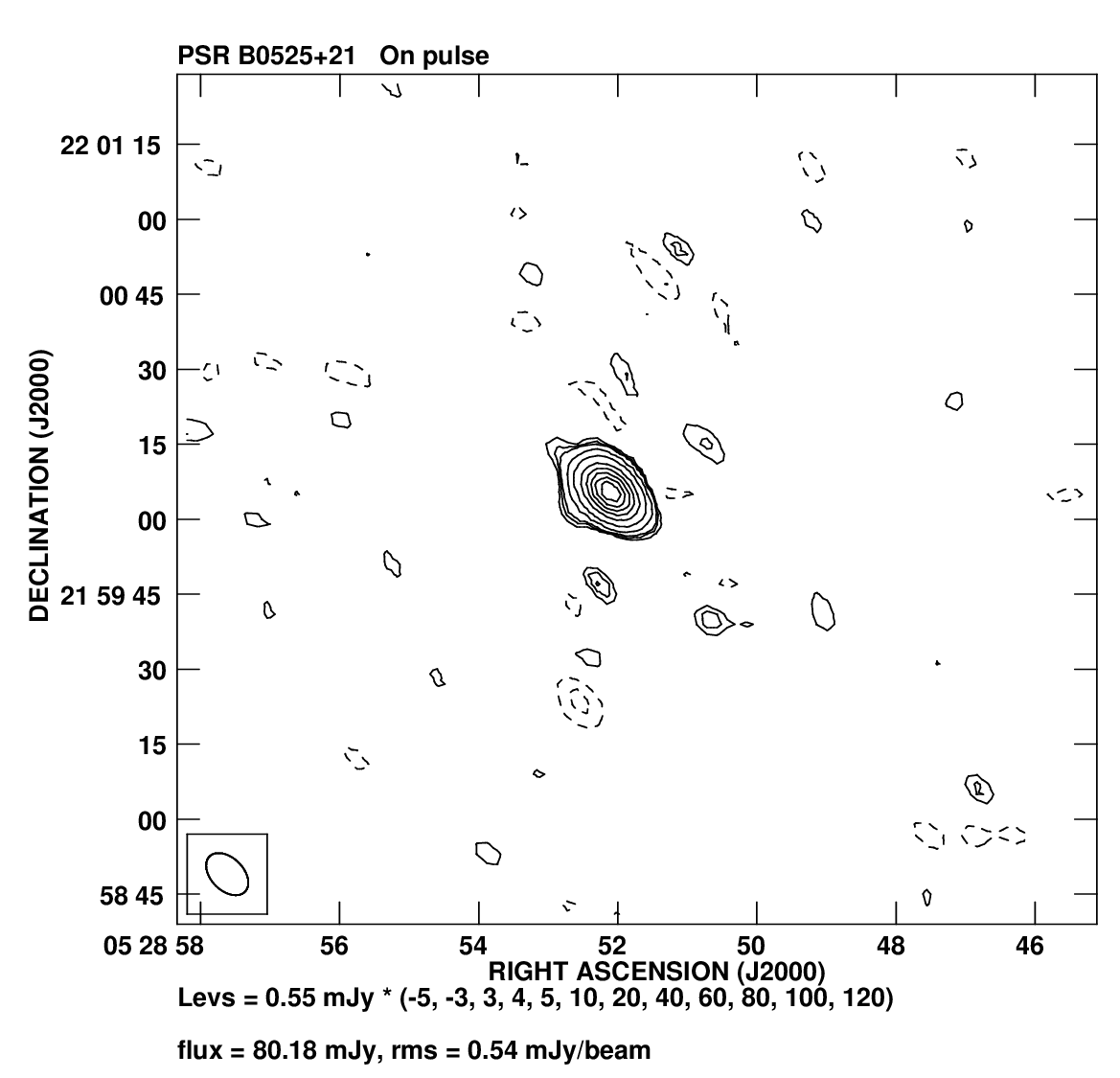}
\includegraphics[angle=-90,scale=.40]{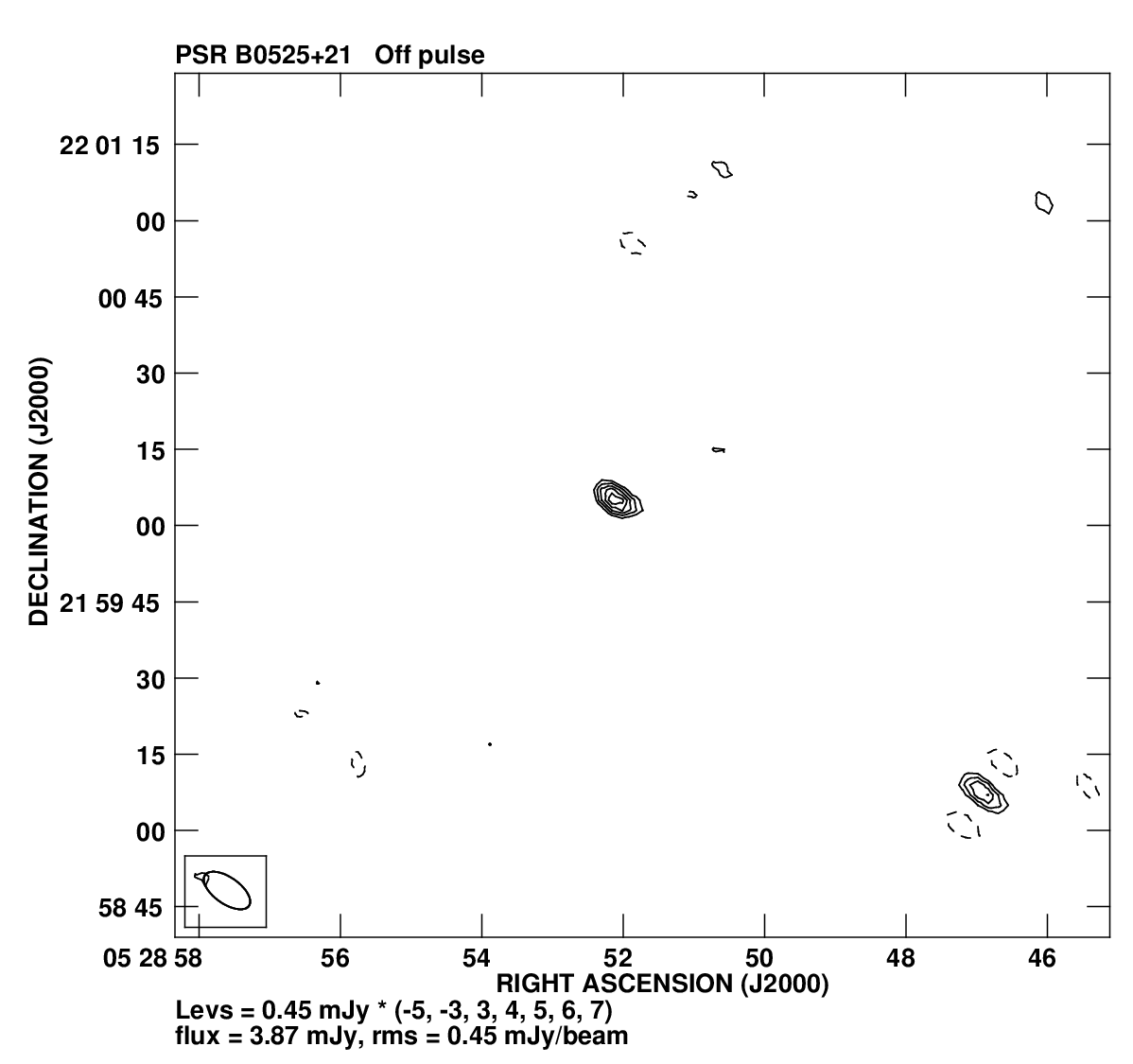}
\caption{Contour plots of the on- and off-pulse images of B0525+21 showing 
the pulsar detection. The synthesized beam is shown in box.
\label{fig_P1imag}}
\end{figure}

\begin{figure}
\includegraphics[angle=0,scale=.40]{Offpaper1_fig3a_CntrP2N.eps}
\includegraphics[angle=-90,scale=.40]{Offpaper1_fig3b_CntrP2F.eps}
\caption{Contour plots of the on- and off-pulse images of B2045--16 showing  
the pulsar detection. The synthesized beam is shown in box.
\label{fig_P2imag}}
\end{figure}

\begin{figure}
\includegraphics[angle=0,scale=.60]{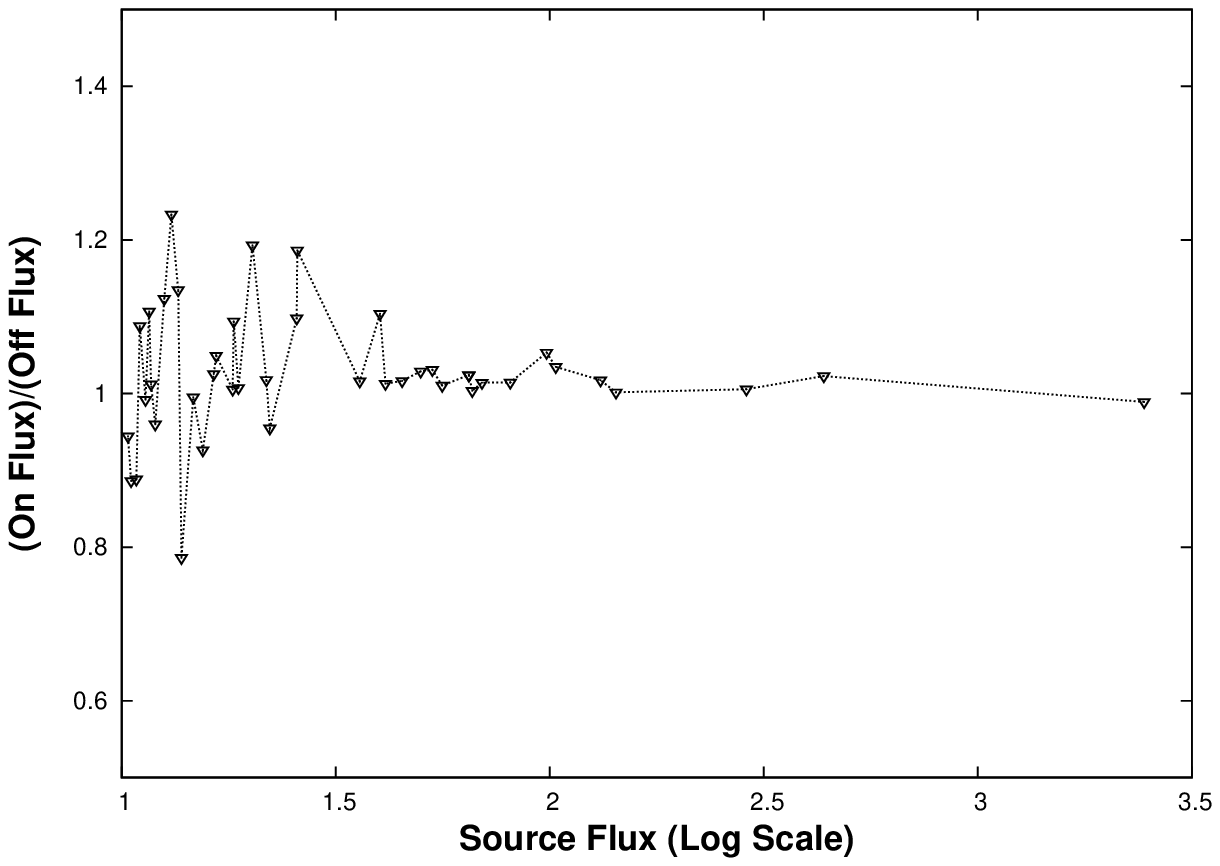}
\includegraphics[angle=0,scale=.60]{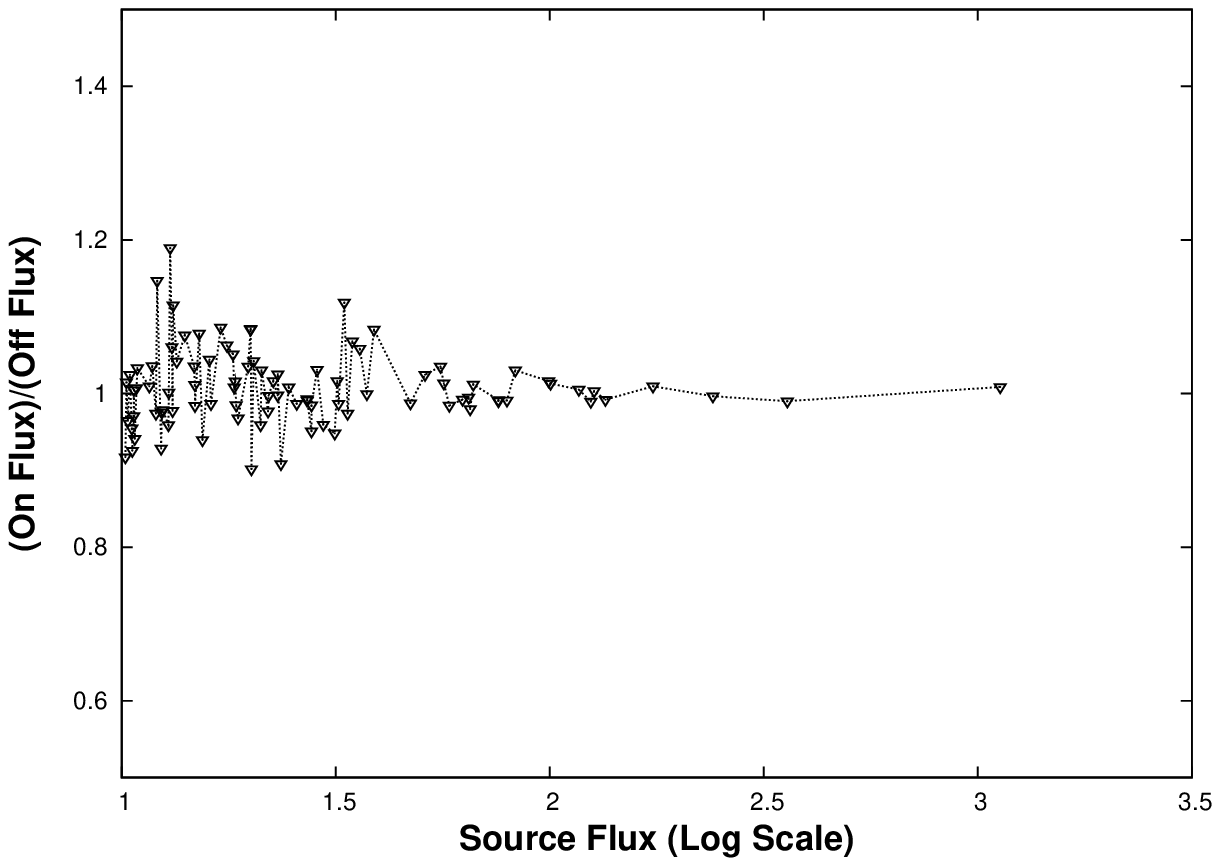}
\caption{The ratio of the measured flux-densities of point sources (flux 
density $>$ 10 mJy) in the on- and off-pulse images. The scatter around unity
confirmed that the flux-scale was the same for both. (B0525+21: top;  
B2045--16: bottom) \label{fig_OnVoff}}
\end{figure}

\section{\large Results}

\begin{table*}
\begin{center}
\caption{Summary of on- and off-pulse observations and results.
\label{Results}}
\begin{tabular}{lccccccccc}
\tableline\tableline
Pulsar &frequency& time &   beam  & size &    On rms   &   On Flux   &
  Off rms    &   Off Flux  &  Int. Flux  \\
       &  (MHz)  & (min)&(\arcsec)& (pc) &(mJy$b^{-1}$)&(mJy$b^{-1}$)&
(mJy$b^{-1}$)&(mJy$b^{-1}$)&(mJy$b^{-1}$)\\
\tableline
B0525+21 & 332.9 & 160 &  9.5$\times$6.5 & 0.088 & 0.55 &  80.2$\pm$5.8 & 0.45 & 3.9$\pm$0.5 & 30.0$\pm$2.1 \\
B2045--16& 317.1 & 180 & 11.9$\times$7.2 & 0.042 & 2.10 & 305$\pm$22& 0.65 & 4.3$\pm$1.1 & 110.5$\pm$7.9\\
\tableline
\end{tabular}
\end{center}
\end{table*}

We detected off-pulse emission in both the pulsars, coincident with the 
location in the on-pulse image. The results are summarized in table 
\ref{Results}. The on- and the off-pulse emissions appeared to be 
unresolved for both the pulsars.

We compared the flux densities of sources in the field of view to check the
flux-density calibration. The ratios of on- and off-image flux densities of
sources stronger than 10 mJy are plotted in figure \ref{fig_OnVoff}. There were
43 and 93 comparison sources in the fields of B0525+21 and B2045--16,
respectively. The scatter of values around unity, especially for strong sources
(which have smaller fractional flux errors) confirmed the similarity of the
flux scale of the on- and off-images for both the pulsars.

{\bf PSR B0525+21:} The pulsar was detected in the off-pulse image at the 
8.6$\sigma$ level (flux-density 3.9$\pm$0.5 mJy, rms noise 0.45 mJy/beam). 
The flux density of the pulsar in the on-pulse image was 80.2$\pm$5.8 mJy 
(rms noise 0.55 mJy/beam). The resolution element at the distance of the 
pulsar corresponds to 0.09 pc. The data from 19 May, 2009, yielded an off-pulse
detection of 3.6 mJy.

{\bf PSR B2045--16:} The pulsar was detected in the off-pulse image at the 
6.6$\sigma$ level (flux density 4.3$\pm$1.1 mJy; rms noise 0.65 mJy/beam). 
The flux density of the pulsar in the on-pulse image was 305.2$\pm$21.9 mJy 
(rms noise 2.1 mJy/beam). The resolution element at the distance of the pulsar 
corresponds to 0.04 pc.

The flux-densities averaged over the entire pulsar period is 30.0$\pm$2.1 for
B0525+21 and 110.5$\pm$7.9 for B2045--16. These values are smaller than
expected (Table \ref{PulsProp}) by factors of 1.5 and 2.7, which can be caused
by effects such as refractive interstellar scintillations \citep{stine90}. We
confirmed the correctness of our flux scale by comparing the flux densities of
the phase calibrators to their known values (less than 1 per cent) and by
comparing the flux densities of 3-5 strong sources in each field to the
interpolated values from NVSS (1.4 GHz) and VLSS (74 MHz).

The quoted errors on the flux densities were obtained by adding in quadrature 
the image rms, calibration errors and the error on the value of the primary 
flux-density calibrator \citep{baars77}.

The signal-to-noise ratio in the published profiles of the pulsars 
\citep{gould98} are insufficient to conclude whether the off-pulse emission 
detected here is a pedestal (throughout the period) or a pulse confined 
to the off-pulse gate.

\section{\large Genuineness of the Off-Pulse Detection}

Firstly, we detected the off-pulse emission in two data sets of B0525+21 
observed 8 months apart, and within the noise the two measurements were 
identical. We can safely rule out the detections being chance occurance 
of noise peaks. Secondly, B2045--16 was positioned almost at the half-power 
point of the primary beam while B0525+21 was close to the field centre. 
The detection in both cases suggests that they are not the spurious structures 
occasionally seen at the centre of the field. We discuss below several effects
which could result in spurious sources in the off-pulse data at the location 
of the pulsar; and demonstrate that they are unlikely to be responsible for 
the same.

\subsection{Positional coincidence of an unrelated source}

To determine the probability of finding an unrelated source coinciding with the
pulsar we used the VLA FIRST catalogue \citep{becker95} to determine the 
following relationship between source counts (N) and flux-density (S):
\begin{equation}
logN = 2.2 - 0.826\times logS + log(\delta S) + log(A_{sky}). 
\label{eqn_logNlogS}
\end{equation}
where $S$ is the flux at 1.4 GHz in mJy, $\delta S$ is the flux bin width at 
1.4 GHz in mJy and $A_{sky}$ is the area of sky under consideration in 
$degree^{2}$.

The area of the synthesized beam is $\sim$100 $arcsec^{2}$ at 325 MHz. The 
flux-density for consideration in the equation is between 3$\times$image-rms 
and 5-10 per cent of the pulsar flux, translated to 1.4 GHz (using spectral 
index $\gamma = 0.75$). The upper value is related to a realistic upper limit 
to the flux fraction in the off-pulse from the single-dish pulse profile 
\citep{gould98}. This yielded a probability of coincident
unrelated sources of $\sim1.5 \times 10^{-3}$. The probability of detecting 
coincident sources for both pulsars is the square of the above number and 
therefore highly improbable.

\subsection{Error in Time stamp}

\begin{figure}
\includegraphics[angle=0,scale=.60]{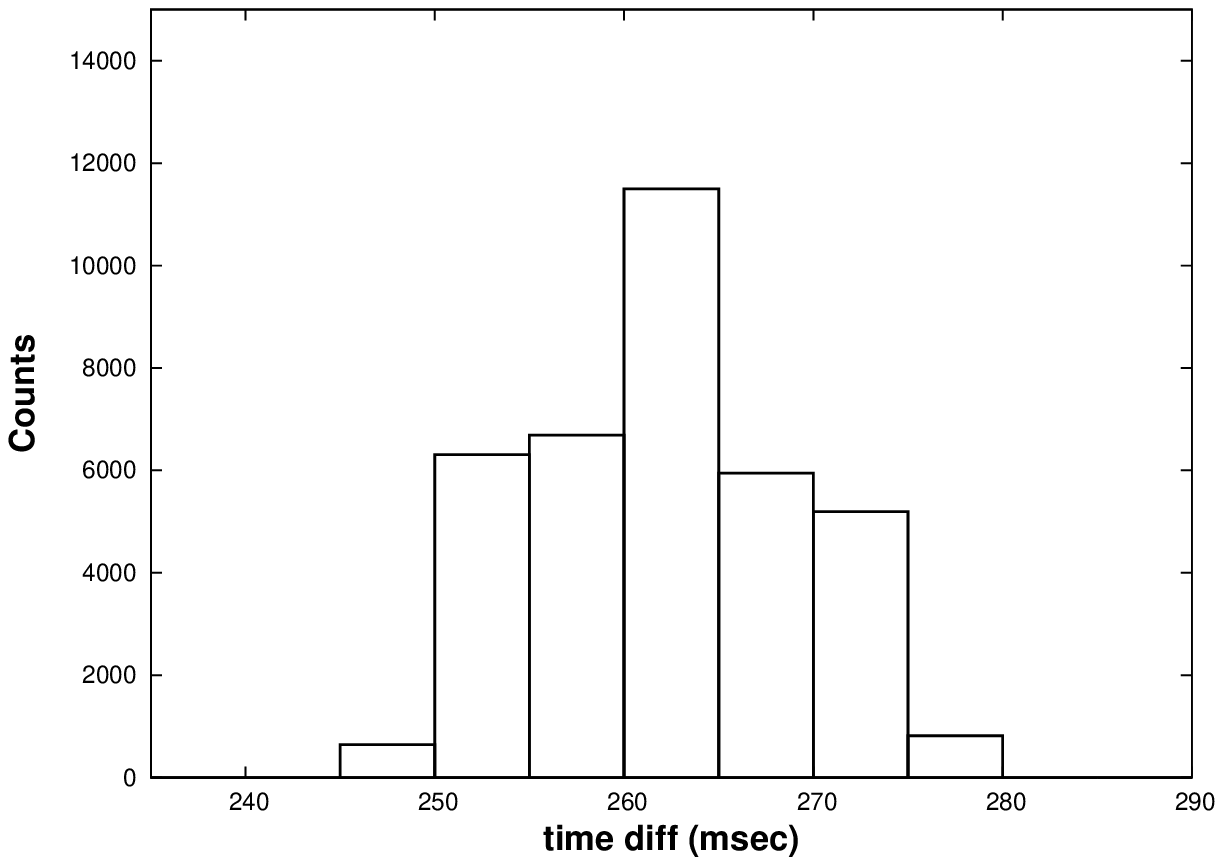}
\includegraphics[angle=0,scale=.60]{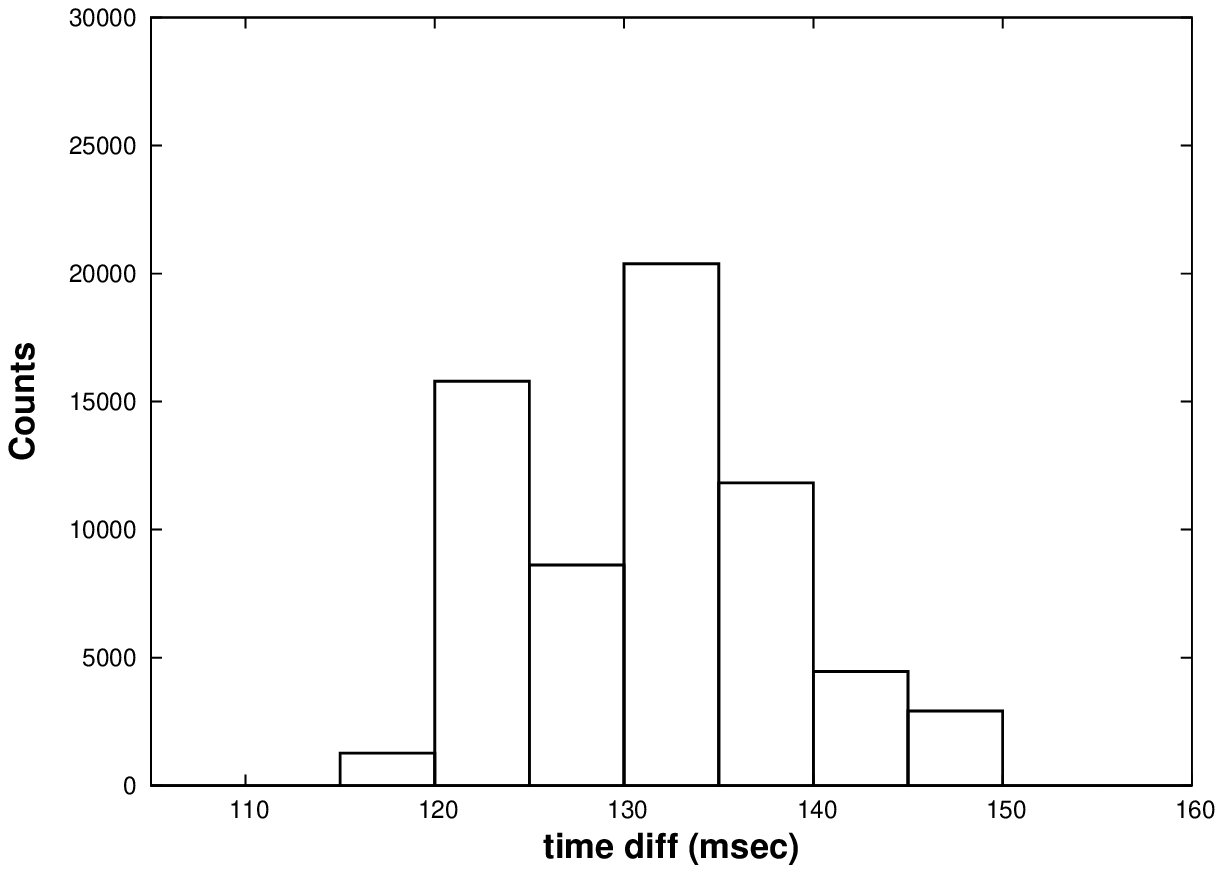}
\caption{Histograms of the time difference between adjacent time-stamps 
for PSRs B0525+21 (top) and B2045--16 (bottom). The rms of the plots is 7 ms 
for both.\label{fig_timehist}}
\end{figure}

This scenario requires that approximately 1-4 per cent of the on-pulse 
data fall within the off-pulse gate. However, any monotonic drift in the 
clock, without recovery, would have smeared out the pulse obtained by folding 
the self-data; on the contrary we see the on-pulse detected with a signal-
to-noise ratio of 50-100 (figure~\ref{fig_profileP}). 
Alternatively, the clock could have maintained an accurate long term average 
but with large excursions in the values of the individual time stamps, i.e. 
excursions of the order of 1-2 s (half the pulsar period) for an output 
data rate of one visibility per 0.131s or 0.262s. This sounds somewhat contrived; 
if this were the case we should have expected a histogram of the time difference 
between adjacent data to show a large scatter. The histograms plotted in 
figure~\ref{fig_timehist} show a tight scatter (rms 0.007s) around the expected
mean values (0.262s and 0.131s).

Additionally, any artifact due to time-smearing should have resulted in the 
shorter-period and faster-sampled B2045--16 having a higher off-pulse fraction,
whereas the ratio is much higher for B0525+21. 

\subsection{Leakage of the signal along the time series}

This scenario requires temporal correlation between data separated by at least 
half the pulsar period. The basic sampling interval of the GMRT correlator is 
31.25 nanoseconds. A cross-spectrum is output every 16$\mu$s by fourier 
transforming 512 such samples. The data output every 16$\mu$s should be
independent from all other such data in the time series. 8188 (or 16376) such
independent samples are averaged to obtain the short-term acquisition (STA)
date output at 0.131 (or 0.262 s).  Given the independence at the 16$\mu$s 
level it is difficult to envisage correlation between one STA and
the next, even more so across 6-10 STAs (half the pulsar period). Nevertheless
this was a possibility. Observers using the standard interferometric mode of
the GMRT would not be affected as none of their targets
vary on sub-second timescales, and their integration time is rarely less than
2s (more typically 17s). The leakage would only redistribute the flux density
along the time axis without causing any change in the measured flux density.
Observers using the pulsar mode would not have detected such an effect because
they are only sensitive to 
the relative flux in excess of the base level.

We tested this possibility by estimating the temporal correlation in the 0.131s
data series. Assuming that the off-pulse was entirely due to leaking of the 
signal from the on-pulse bin to the off-pulse bins we estimated the required 
leakage for a spurious detection of the observed level. There was no way of
introducing a narrow, clean pulse into the GMRT receiver system to directly 
measure such a leakage. So, we investigated the consequence of such a 
correlation on the noise data.

First, we ran simulations to estimate the auto-correlation function for noise 
in the presence of leakage sufficient to generate the observed off-pulse, i.e. 
we generated a noise series and smeared each data point into subsequent data 
points according to a particular time profile --- e.g. constant leakage into the
next 13 bins; or linearly decreasing leakage into the next 13 bins. The value 
13 is related to the period of the two pulsars which covered 14 time bins.

We then recorded 6$\times$10 minute scans of noise data (0.131s integration) 
but with Front-End termination which sealed the telescope at the antenna feeds.
This ensured that no temporally continuous external source (cosmic sources and 
RFI), which would be correlated across the entire observing session, was 
present in the noise data.

\begin{figure}
\includegraphics[angle=0,scale=.60]{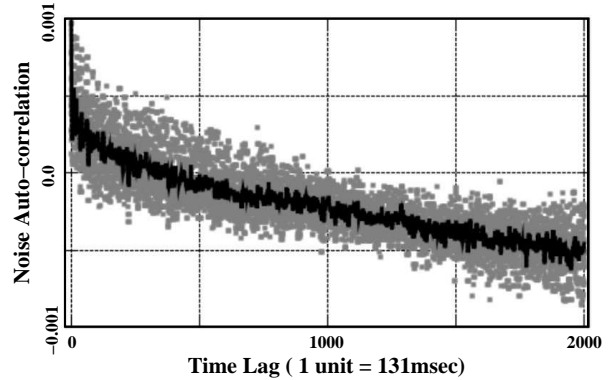}
\caption{Auto correlation profile for front-end terminated GMRT noise data. 
The scatter plot is for each baseline-channel-scan while the black line is the
median profile from the scatter. The non-zero value even at large lags is 
indicative of non-stationary noise signal and system gain variation.
\label{autocorrFull}}
\end{figure}

The auto-correlation for each baseline-channel (435$\times$120) data was
separately calculated for each scan. The median autocorrelation profile and the
scatter are shown in Figure~\ref{autocorrFull}. The worst data are at the 0.1
per cent level but the median is only about 0.04 per cent. The profile does not
fall to zero at large time lags, indicating non-stationarity of the noise
signal. We obtained a similar behaviour by introducing a time-varying mean
level into the noise in our simulations. This time variation in the mean level
may reflect the system gain variations as well, which would have been corrected
to a greater or lesser extent by self-calibration during imaging. Therefore the
profile is a firm upper limit to the contribution of temporal leaking of signals
to an off-pulse detection.  

It should be noted that interferometric imaging is an excellent filter of bad 
data; a compact source (like the off-pulse detection) would require most of the 
baselines to show high temporal auto-correlation; a few bad baselines with high 
auto correlation would only result in noisy ripples across the image and not a
localised source. Therefore the median line is a more accurate measure of the
leakage for our study than the extreme points in the scatter. 

We note that non-stationarity and gain fluctuation only occur at the 0.05 per 
cent level in the GMRT, and would not have a discernible effect for most 
imaging exercises. 

\begin{figure}
\includegraphics[angle=0,scale=.60]{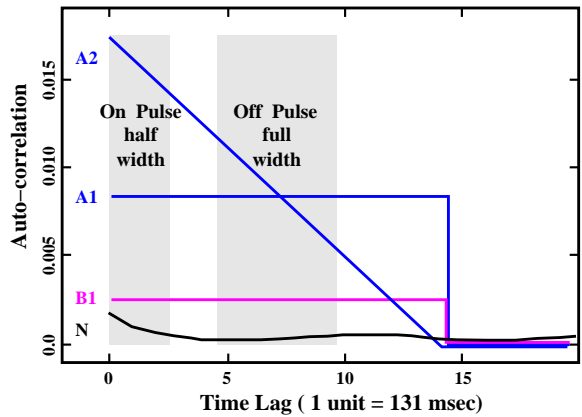}
\caption{An expanded view of the short-lag regions of the auto-correlation 
profile plotted in Figure~\ref{autocorrFull}. The plots A1, A2 (PSR B0525+21) 
and B1 (PSR B2045--16) are the required (simulated) noise auto-correlation for 
a spurious off-pulse detection due to temporal leaking of the pulse into 
subsequent bins. The thick black curve is the observed noise auto-correlation 
profile (from Figure~ \ref{autocorrFull}). \label{autocorrMag}}
\end{figure}

Figure~\ref{autocorrMag} shows a magnified representation of the relevant 
part of the noise auto-correlation plot. A1 and A2 (both PSR B0525+21) and 
B1 (PSR B2045--16) indicate the required noise auto-correlation if the detected 
off-pulse were due to the leaking of the signal from the on-pulse bin.
The analysis recipe for PSR B0525+21 (with similar arguments for PSR B2045--16) 
is as follows:\\
1. The pulsar period includes 14 time bins, each of 0.262 s. The on- and 
off-pulse fluxes measured over 5 time bins are 80.2 and 3.9 mJy, respectively.
Higher time resolution observations \citep{gould98} show that the main-pulse 
narrower than one of our bins.\\
2. A leakage of 0.9 per cent of the pulse flux from the first bin into each 
of the next 13 bins will explain the observed off-pulse. This corresponds to an
auto correlation level of 0.0087. The auto-correlation profile calculated for 
GMRT noise (the lowest curve 'N' in figure~\ref{autocorrMag}), is smaller than 
the required amount by a factor of at least 20.\\
3. A leakage reducing with temporal distance is more likely. A linearly 
reducing leakage (curve A2) requires a similar average auto-correlation in the 
off-pulse region and a higher correlation at short lags.\\
4. We conclude that temporal leaking of the signal is incapable of explaining 
the detected level of off-pulse emission.

The curve B2 represents the corresponding model for PSR B2045--16: off-pulse 
4.3 mJy; on-pulse flux 305.2 mJy; required constant leakage 0.43 per cent; 
required correlation in the off-pulse region 0.003; discrepancy is a factor 
of $\sim$ 6.

Additionally, the argument against time-smearing (Sec. 4.2, second para) is 
also valid against leaking of the signal along the time series.

\section{\large Discussion}

\subsection{Previous Searches for Off-Pulse Emission, including PWNe}

A literature survey yielded many reports of searches for ``continuous emission 
in the direction of pulsars'' (Table \ref{LitrSurvey}). Here we shall discuss 
only those in which gating techniques were applied to target off-pulse 
emission, which could have been either from the pulsar itself or from PWNe. 

Perry \& Lyne (1985) reported off-pulse emission from 4 pulsars (PSRs B1541+09,
B1604--00, B1929+10 and B2016+28) using a gated two-element interferometer 
at 408 MHz. Subsequent studies showed that two of these pulsars, B1541+09 and 
B1229+10, were aligned rotators, undermining the claim of off-pulse 
origin \citep{hank93, rat95}. The off-pulse emission in the other two pulsars, 
B1604-00 and B2016+28, were later identified with unrelated background sources
(Storm \& Van Someren Greve 1990; Hankins et al. 1993).

Bartel et al. (1985) reported a non-detection from PSR B0329+54 using gated 
VLBI interferometry at 2.3 GHz. Strom \& Van Someren Greve (1990) reported 
non-detections in 43 pulsars using the WSRT in gated mode at 327 and 609 MHz. 
Stappers et al. (1999) reported non-detection in 4 pulsars using ATCA in 
gated interferometric mode at 1.3 and 2.2 GHz.

Gaensler et al. (2000) searched for unpulsed emission from 27 energetic and/or 
high velocity pulsars using gating interferometry with the VLA and ATCA at 
1.4 GHz and found emission in two cases. However they could not conclusively 
determine whether these detections corresponded to PWNe or the pulsar 
magnetosphere; they even suggested that the detections could be spurious and 
a result of improper online gating. These detections were at the level of
2$\sigma$ (PSR B1634-45) and 13$\sigma$ (PSR B1706-16).

There have been detections of several radio pulsars with associated plerionic 
bow shock nebula: B1951+32 \citep{hester98}, B1757--24 \citep{frail91}, 
B1853+01 \citep{frail96} and B1643--43 \citep{giacani01}. All of these are 
highly energetic pulsars associated with SNR \citep{chat02}. Gaensler et al. 
(1998) reported the detection of off-pulse emission from PSR B0906--49, which 
appears to be a PWN generated by a slow moving pulsar in the dense ISM.


\begin{table*}
\begin{center}
\scriptsize
\caption{Literature survey for off-pulse studies.\label{LitrSurvey}}
\begin{tabular}{llll}
\tableline\tableline
Non detections/   &    Detections    &  Tentative  &  PWNe detection \\
subsequent SNR    &  later refuted   &  detection  &                 \\
\tableline
1. Sch\"{o}nhardt 1973, 1974 &1. Gopal-Krishna 1978 &1. Gaensler et al. 2000 ($G$)&1. Frail \& Kulkarni 1991 \\
2. Weiler et al. 1974        &2. Glushak et al. 1981&                       &
2. Frail et al. 1996      \\
3. Cohen et al. 1983         &3. Perry \& Lyne 1985 ($G$)&                  &
3. Gaensler et al. 1998    \\
4. Bartel et al. 1985 ($G$)  &                      &                       &
4. Giacani et al. 2001    \\
5. Strom \& Van Someren      &                      &                       &
                          \\
   Greve 1990 ($G$)          &                      &                       &
                          \\
6. Hankins et al. 1993 ($G$) &                      &                       &
                          \\
7. Frail \&                  &                      &                       &
                          \\
  Scharringhausen 1997       &                      &                       &
                          \\
8. Stappers et al. 1999 ($G$)&                      &                       &
                          \\
                          \\
\tableline
\end{tabular}
\tablenotetext{~}{There have been no unambiguous detection of off-pulse 
emission not associated with PWNe. The gated interferometry studies in table~
\ref{LitrSurvey} have been indicated with $G$.}
\end{center}
\end{table*}


\subsection{Our detections in the context of previous effort}

We discuss here the unsuccessful search for off-pulse emission in PSR B0525+21 
by Weiler et al. (1974) and Perry \& Lyne (1985). Weiler et al. (1974) were 
looking for extended emission around the pulsar without employing any kind 
of gating. They had a synthesized beam size of 47\arcsec $\times$ 133\arcsec 
with a noise rms of 2.5 mJy/b. It is apparent that their setup was inadequate 
for the present level of detection. Perry \& Lyne (1985) employed a two-element 
interferometer with hardware gating at 408 MHz for their studies but could 
neither image the fields and nor correct for sensitivity variations. They
listed considerable emission in the off-pulse bin for B0525+21 (in fact 
greater than for B1929+10 and B2016+28, which were claimed to be positive 
detections) but the noise was also very high precluding a positive detection.

The major advantages of our effort were:\\
1. Low frequency of observation: the steep spectrum pulsar is likely to 
dominate over emission from any associated nebula with flatter spectrum;\\
2. A multi-element interferometer: The GMRT is currently the most sensitive 
low-frequency instrument. The ability to correct gain variations, filter
bad data using imaging residuals and RFI excision tools \citep{athreya09} 
allowed us to make sufficiently deep images.\\
3. Off-line gating of data: this allowed secure separation of the off- 
and on-pulse sections.

\subsection{PWN emission}

The confinement of the relativistic wind from pulsars generates PWNe
which are luminous across the electromagnetic spectrum in synchrotron,
inverse compton, and optical line emission from the shocked regions.
All known PWNe seen around radio pulsars have spin-down luminosities
$\dot{E} \gtrsim 10^{34}$~erg~s$^{-1}$. They all appear to be young and, with 
the exception of PSR B0906--49 \citep{gaens98}, are associated with SNR. Two 
classes of PWNe can be formed in the absence of associated SNR: one,
where the pulsar wind is confined by the density of the ISM --- known as 
static PWN; and the other, where the wind is confined by the ram-pressure of
motion of the pulsar through the ISM --- called bow-shock nebula. Our two 
target pulsars, B0525+21 and B2045--16, are not associated with SNRs. So, we 
explored the possibility that the detected off-pulse emission was due to a PWN.

The efficiency factor $\epsilon_{R}$ is defined as the ratio of the radio 
bolometric luminosity ($L_{R}$) of a PWN to the spin-down energy ($\dot{E}$)
of the pulsar $L_{R} = \epsilon_{R} \dot{E}$. If we assume a typical
PWN spectral index of $\gamma \sim$ 0.3 its radio luminosity between $10^{7}$
Hz and $10^{11}$ Hz is given by $
L_{R} = 3.06 \times 10^{28}~d_{kpc}^{2} S_{mJy} ~ {\rm erg ~ s^{-1}},
$
where $d_{kpc}$ is the distance to the PWN in kpc and $S_{mJy}$ is 
the integrated flux of PWN at 325 MHz in mJy. Using the measured 
flux-density (table~\ref{Results}), source distance and the spin-down energies
(table~\ref{PulsProp}) we calculate efficiency factors of 
2$\times 10^{-2}$ for B0525+21 and 2$\times 10^{-3}$ for B2045--16. These 
$\epsilon_{R}$ values are 1-2 orders of magnitude higher than those of 
previously known PWNe (typical $\epsilon_{R} \sim 10^{-4}$; see Frail \& 
Scharringhausen 1997, Gaensler et al. 2000 for a discussion).

\paragraph{Static PWN} The relativistic particles and 
Poynting flux emanating from the pulsar, at rest relative to the ISM, will 
drive through the ambient medium a shock of radius $R_{s}$ given by 
\citep{arons83}
\begin{equation}
R_{s} = \left(\frac{\dot{E}}{4\pi\rho_{o}}\right)^{1/5}~{t^{3/5}}
\label{eqn_rs}
\end{equation}
The velocity of the shock front is given by
\begin{equation}
\dot{R}_{s} = \frac{3}{5}\left(\frac{\dot{E}}{4\pi\rho_{o}t^{2}}\right)^{1/5}
=3.3 \left(\frac{\dot{E}_{31}}{t_{6}^{2} n_{0.01}}\right)^{1/5}~{\rm km~s^{-1}}
\label{eqn_rsv}
\end{equation}
Here $\rho_{o} = m_{H}n$, where $m_{H}$ is the proton mass and $n$ 
is the particle density of the ISM. Using equation (\ref{eqn_rs}) the required
particle density for a PWN is
\begin{equation}
n = 5.35 \times 10^{11} \left(\frac{\dot{E}_{31}t_{6}^{3}}{R_{0.01}^{5}}\right)~ {\rm cm^{-3}}
\label{eqn_nrs}
\end{equation}
where $\dot{E}_{31}$ is the spindown power in units of $10^{31}$~erg~s$^{-1}$,
$t_{6}$ is the age in units of $10^{6}$ yr, $R_{0.01}$ is the radius of PWN 
in units of 0.01~pc and $n_{0.01}$ is the ISM density in units of 0.01 ${\rm
cm^{-3}}$. In the present exercise we assume that the PWNe in our targets are 
smaller than one synthesized beam width since the off-pulse emission is 
unresolved in both pulsars. Using equation (\ref{eqn_nrs})
and values in tables \ref{PulsProp} and \ref{Results} we determined the ISM 
density required to drive a PWN to be $\sim3 \times 10^{9}~{\rm cm^{-3}}$ for 
B0525+21 and $\sim10^{12}~{\rm cm^{-3}}$ for B2045--16. The required particle 
densities are much higher than typical ISM densities of 
$\sim0.03~{\rm cm^{-3}}$, suggesting that these pulsars are too weak and old to 
power a static PWN through the ISM. Since our estimate of the sizes of the 
nebulae are upper limits the particle density estimates are lower limits.

\paragraph{Bow Shock PWN} This requires that the shock velocity ($\dot{R}_{s}$) 
be much smaller than the pulsar transverse velocity ($V_{PSR}$). The typical 
ISM density of $0.03 {\rm cm^ {-3}}$ in equation~(\ref{eqn_rsv}) yields 
$\dot{R}_{s}$ $\sim 2.8~{\rm km~s^{-1}}$ for B0525+21 and 
$\sim2.5~{\rm km~s^{-1}}$ for B2045--16, which are indeed much smaller than the 
$V_{PSR}$ listed in Table \ref{PulsProp}. The radius of the shock is given by 
\citep{frail97}
\begin{equation}
R_{BS} = \left(\frac{\dot{E}}{4\pi c \rho_{o} V_{PSR}^2}\right)^{1/2}
\label{eqn_rbs}
\end{equation}

From equation (\ref{eqn_rbs}) the particle density required to sustain a 
bow shock PWN is 
\begin{equation}
n = 1.67 \times 10^{-4} \left(\frac{\dot{E}_{31}}{R_{0.01}^{2}V_{100}^{2}}
\right) {\rm cm^{-3}}
\label{eqn_nrbs}
\end{equation}
where $\dot{E}_{31}$ is the spin-down power in units of $10^{31}~{\rm erg~s^
{-1}}$, $R_{0.01}$ is the radius of PWN in units of 0.01~pc and $V_{100}$ is 
the velocity of the pulsar through the ISM in units of 100 ${\rm km~s^{-1}}$.
This requires an ISM density of $\sim5\times10^{-6}{\rm cm^{-3}}$ for 
B0525+21 and $\sim8\times10^{-6}~{\rm cm^{-3}}$ for B2045--16. These values 
are 3-4 orders of magnitude lower than the typical ISM density. This implies 
that the pulsars are too weak to drive a bow shock nebula of the size 
corresponding to the telescope resolution. However, a bow shock interpretation 
can be salvaged if the nebulae are $\sim$2 orders of magnitude smaller.

In summary, identifying the off-pulse emission detected here with PWNe results
in unrealistic values of the ISM particle density. This is not surprising 
because we selected as our targets middle-aged pulsars, which category has 
hitherto not been known to be associated with PWNe.

\subsection{Emission from the Magnetosphere}

Numerous published studies provide estimates of $\alpha$ and $\beta$ for the 
pulsars studied here. Some use an empirical/geometrical (E/G) approach to 
establish $\alpha$ and $\beta$ (Lyne \& Manchester 1988; Rankin 1993; see 
Everett \& Weisberg 2001 for a discussion on the E/G approach). One can also
obtain $\alpha$ and $\beta$ values by fitting the RVM to the PPA traverse
\citep{mitra04}, but these fits yield highly correlated estimates of the two
angles. This issue has been discussed in depth by Everett \& Weisberg (2001) 
and Mitra \& Li (2004).

Everett \& Weisberg (2001) list six different studies of PSR B0525+21 with
$\alpha$ ranging from 116\degr\ to 159\degr\footnote{Following 
Everett \& Weisberg (2001) $\alpha$ values should span 0-180\degr (and not 
0-90\degr) for a consistent definition of the pulsar rotation axis. 
Everett \& Weisberg provide a compilation of published $\alpha$ and $\beta$ 
values corrected for this convention.}. Mitra \& Li (2004) estimated $\alpha 
\sim 127-144\degr$ for PSR B2045--16. Both RVM fit and the E/G approach
were used to obtain the above estimates of $\alpha$. These studies provide
good evidence that our targets, B0525+21 and B2045--16, are not aligned rotators
and therefore any off-pulse emission must arise far from the magnetic pole.

The pulse profiles of both these pulsars show sharp rising and falling edges. 
PSR B0525+21 is an example of a classical double (D) profile, whereas PSR 
B2045--16 is classified as a triple \citep{rankin93}. The main-pulse width for 
these pulsars span about 5 per cent of the period, and there are no 
obvious emission components visible outside the main-pulse. Based on 
the pulse widths and the estimates of $\alpha$ and $\beta$, Mitra \& Li (2004) 
concluded that the radio emission arises at a distance of 1--2 per cent of the 
light cylinder. Hence, the radio on-pulse emission of PSR B0525+21 and B2045–16
are classical examples of emission from open dipolar field lines. The remaining
95 per cent of the period, i.e. the off-pulse region, were hitherto thought to
be radio quiet zones of the pulsar magnetosphere \citep{gold69, rud75}.

Our off-pulse phase corresponds to 80-208\degr (from the peak of the main-pulse)
for B0525+21 and 101-229\degr for B2045--16. In a few cases pre-/post-cursor
emissions have been detected about 60\degr from the main-pulse, posing questions
regarding their origin \citep{mitra10}. Our detections are even further away 
from the main-pulse. It remains to be seen if these off-pulse detections are an
extreme example of PPC  or if it represents unpulsed emission from throughout 
the pulse period.


It is to be noted that the opening angle of dipolar field lines scales as the 
square of the emission height. If radio emission from pulsars originate from 
higher up the light cylinder ($\gg$ 1-2 per cent of the light cylinder), the 
polar emission could span a larger longitudinal range. This implies that radio 
emission at larger distances from the neutron star surface are potential sites 
of PPC or off-pulse emission in pulsars. Alternatively, if the 
emission arises close to the neutron star surface, the emitting region will
need to encroach upon the closed field lines of the neutron star. Thus, the 
existence of magnetospheric off-pulse emission should prove to be an important 
input for pulsar electrodynamic models which try to establish the relation 
between magnetospheric currents and coherent pulsar radio emission 
\citep{spit06}.

\section{\large Conclusion}

We report GMRT observations at 325 MHz which detected off-pulse emission from 
two long-period pulsars which have low spin-down rates and are not associated 
with supernova remnants. We have adduced evidence that the signals are not 
artifacts of the observing procedure, nor unrelated background sources which 
happen to be coincident with the pulsar. We have argued that explaining these 
detections as Pulsar Wind Nebula requires ISM particle densities which differ 
from typical measured values by several orders of magnitude. Robust estimates 
of the geometrical parameters of these two pulsars argue against them being 
aligned rotators. This leaves the possibility of emission from the 
magnetosphere. If the off-pulse emission arises from a much higher height 
than the on-pulse then it could arise from open field lines at the edge of 
the light cylinder. On the other hand if the off-pulse emission is from the 
same height as the on-pulse emission then it must be associated with closed
field lines. Further studies at multiple frequencies, in polarization and 
higher resolution are needed to establish the nature of this emission which 
can impose valuable constraints on the pulsar emission mechanism models.

\acknowledgments
We thank the director of NCRA for discretionary observing time which made this
work possible. We also thank the staff of the GMRT for their help with these
non-standard observations possible. The GMRT is run by the National Centre for
Radio Astrophysics of the Tata Institute of Fundamental Research. We also
thank the referee, Scott Ransom, for useful comments.

{\it Facilities:} \facility{GMRT}.

\clearpage

\end{document}